# Equilibrium Time, Permutation, Multiscale and Modified Multiscale Entropies for Low-High Infection Level Intracellular Viral Reaction Kinetics

Fariborz Taherkhani[1], Farid Taherkhani [2*]

[1]*Department of Computer Science, University of Wisconsin-Milwaukee, Milwaukee, WI, USA*

[2] *Departments of Chemistry Sharif University of Technology, Tehran, Iran*

Corresponding Author:

faridtaherkhani@gmail.com

## Abstract

Viral infectious diseases, such as HIV virus growth, cause an important health concern. Study of intracellular viral processes can provide us to develop drug and understanding the drug dose to decrease the HIV virus in during growth. Kinetics Monte Carlo simulation has been done for solving Master equation about dynamics of intracellular viral reaction kinetics. Scaling relationship between equilibrium time and initial population of template has been found as power low, $f_{eq\,time}(N) = aN^b$, where $N$, $f_{eq\,time}(N)$ are the number of initial population of template species, equilibrium time, $a = 163.1$, $b = -0.1429$ respectively. Stochastic dynamics shows that increasing initial population of template decreases the time of equilibrium. Entropy generation has been considered in low, intermediate and high infection level of intracellular viral kinetics reaction in during dynamical process. Permutation, multi scaling and modified multiscaling entropies have been calculated for three kinds of species in intracellular reaction dynamics, genome, structural protein, and template. Our result shows that presence of noise in dynamical process of intracellular reaction will change order of permutation entropy for the mentioned of three species. In addition to multiscaling entropy is computed for mentioned model and it has the following order: template > structural protein> genome. Dependency of permutation entropy result to permutation order becomes small in high infection level in intracellular viral kinetics dynamics. At short time scale in intracellular reaction dynamics, convergency of permutation entropy occurs with medium permutation order value. In the big time scale of intracellular dynamics, permutation entropy $H(n)$ scale with permutation order $n$ as a scaling relation $H(n) = n^\alpha$ ($\alpha = 0.30$). Three different kinds of trend for low, medium and high infection level observed for multiscaling entropy of template species versus scaling factor. In high infection level behavior of multiscaling entropy of template is non-monotonic however monotonic behavior of permutation entropy of template can be observed in low and very high infection level. Both multiscaling entropy and modified multiscaling entropy shows one maximum peak as a function of initial template population however permutation entropy versus initial population does not show such a peak and permutation entropy increase with initial template population monotonically.

# 1. Introduction

Various measures of complexity were developed to compare time series and distinguish regular, chaotic, and random variable. The main types of complexity parameter are entropies, fractal dimensions, and Lyapunov exponent[1, 2].

Many ingenious algorithms, tricks, and recipes have been developed during the last 20 years in order to estimate complexity measures from real- world time series. Another way of complexity measure which is used for any time series (regular, chaotic, noisy) is permutation entropy[3]. During the last two decades, a number of interesting methods have been proposed to detect dynamical changes. They include, among others, recurrence plots and recurrence quantification analysis, recurrence time statistics based approaches, space- time separation plots, and their associated probability distribution, statistical test using desterilized invariant distribution in the reconstructed phase space, cross correlation sum analysis, and nonlinear cross prediction analysis[4]. Most of these methods are based on quantifying certain aspect of the nearest neighbors in phase space and as a result are computationally expensive. Recently, Bandt, and Pompe introduced the interesting concept of permutation entropy, as a complexity measure for time series analysis[3].

Permutation entropy for analyzing of time series of heart interbeat signal has been applied for recognition healthy and heart failure people[5].

Diseased systems, when associated with the emergence of more regular behavior, show reduced entropy values compared to the dynamics of free-running healthy systems[5]. However, certain pathologies, including cardiac arrhythmias like atrial fibrillation, are associated with highly erratic fluctuations with statistical properties resembling uncorrelated noise[6- 8]. Viruses are ideal systems for probing how the static linear information encoded by a genome defines a dynamic nonlinear process of growth and development.

The current interest in computational cell biology reflects the widespread belief that the complexity and sophistication of computers and programming could potentially match the complexity of living cells[8].

Our aim of work is understanding of complexity of dynamics of intracellular dynamics help us to predict the behavior of viruses in during of growth process. Entropy value in during time evolution for dynamical of complex bio-chemical reaction will be explored. Natural complexity of intracellular reaction dynamics in low-high infection level will be examined with different entropy approach and ability of them for recognition dynamical complexity. Initial population of template effect on entropy trend in during dynamical process will be explored. We will investigate equilibrium time as function number of initial population of template.

## 2. Intracellular Viral Kinetics

The Viral nucleic acids were classified as genomic (gen) or template (temp).The genome may be DNA and RNA which are positive- strand, negative strand or some other type. According to equations (1-5) we have two ways for changing the genome. The first way, the genome converts to template which has catalytic role for synthesizing every Viral component[9]. The second way it package within structural proteins to form progeny virus with different probabilities. It is very important to notice that the synthesis of structural protein requires that the Viral DNA be transcribed to mRNA and mRNA must be translated to generate the structural protein. At first when the population of template is very high structural protein is synthesized for incorporation into progeny particle. We can show the mechanism of viral kinetics according following reaction:[9]

$$A \underset{k_3}{\overset{k_1}{\rightleftharpoons}} X \tag{1}$$

$$X \xrightarrow{k_2} D \tag{2}$$

$$X \xrightarrow{k_5} B \tag{3}$$

$$B \xrightarrow{k_6} E \tag{4}$$

$$A + B \xrightarrow{k_4} C \tag{5}$$

In this mechanism A, B, C, D, E and X are genome, structural protein, virus, degradation template, degradation or secretion, and template, respectively.

The rate constant and initial molecular populations are taken from reference [9]

| Rate Constant | $k_1 = .025$ day$^{-1}$ | $k_2 = .25$ day$^{-1}$ | $k_3 = 1$ day$^{-1}$ | $k_4 = 0.0000075$ molecules$^{-1}$ day$^{-1}$ | $k_5 = 1000$ day$^{-1}$ | $k_6 = 1.99$ day$^{-1}$ |
|---|---|---|---|---|---|---|
| Initial population | $A = 10$ | $B=0$ | $C=0$ | $D=0$ | $E=0$ | $X=1$ |

Rate constant in dynamical system in random media is a function of time and rate constant can be related to the Fisher information and increasing the reaction order can increase or decrease dynamical disorder [10].

## 2. Theoretical background

### 2.1. Stochastic Algorithm and Simulation

Another way to investigate the kinetics of a small system is stochastic simulation. Up to now several authors have applied the stochastic algorithms [11]. In recent years stochastic modeling has been emerged as a physically more realistic alternative for modeling of the vivo reactions[11-13]. Stochastic and random diffusion model can be described for autocatalytic reaction model such as Lotka-Volterra reaction-diffusion or predator-prey system [14]. Let $X$ be the time of the event. By a constant hazard we mean that:

$$P(X \in (t, t+dt) \mid X > t) = \alpha dt \tag{6}$$

where $\alpha > 0$ is a constant whose value may be calculated as, $\alpha = \sum_{i=1}^{M} W_{j,n} = \sum_{i=1}^{M} \alpha_i$, where

$\alpha_i = W_{j,n} = k_i \dfrac{j!}{n!(j-n)!}$, $k$ is a rate constant. For a small $\delta t$ we will have:

$$P(X \in (t, t+dt) \mid X > t) = \alpha \delta t \tag{7}$$

Considering a time $t > 0$, and a large integer $N$, dividing the interval $(0, t]$ into $N$

subintervals of the form $((i-1)\delta t, i\delta t]$, i=1,2,...,$N$ where $\delta t = \dfrac{t}{N}$, then we have:

$$P(X > t) = P[(X \notin (0, t])] = P(\{(X \notin (0, \delta t])\} \bigcap \{X \notin (\delta t, 2\delta t]\} \bigcap ... \{X \notin (N-1)\delta t, t]\}) \tag{8}$$

Hence

$$\begin{aligned} P(X > t) &= P(X \notin (0, \delta t]) P(X \notin (\delta t, 2\delta t] \mid X > \delta t) \\ &... P(X \notin ((N-1)\delta t, t] \mid X > (N-1)\delta t) \\ &\approx (1 - \alpha \delta t) \times (1 - \alpha \delta t) \times ... (1 - \alpha \delta t) = \\ &(1 - \alpha \delta t)^N \\ &= (1 - \frac{\alpha t}{N})^N \end{aligned} \tag{9}$$

If $N \to \infty$ and $\delta t \to 0$ therefore equation 9 will convert to $\exp(-\alpha t)$ then

$P(X \le t) = (1 - \exp(-\alpha t))$. Consequently, whenever we consider a time dependent event with constant hazard $\alpha$, in Gillespie algorithm[11]. We can conclude that the time distribution is an exponential function. By choosing two uniform random numbers and within the interval [0,1] and by definition of two following expressions, we may write:

$$\tau = \frac{1}{\alpha} \ln(\frac{1}{r_1})$$ (10)

$$\sum_{i=1}^{\mu-1} \alpha_i < r_2 \alpha \le \sum_{i=1}^{\mu} \alpha_i$$ (11)

There are three loops for algorithm as follows:

1) Calculating $\alpha_i = k_i \frac{j!}{n!(j-n)!}$, whereas $k$ is a rate constant. For $1 \le \mu \le M$

2) Generating two uniform random numbers $r_1$ and $r_2$ and calculating $\tau$ and $\mu$ according to equations 10 and 11.

3) Increasing $t$ by $\tau$ and adjusting population of reactants for reaction $\mu$ [11]

## 2.2. Permutation Entropy

Permutation entropy for time series as ($x$ ($i$), $i$=1, 2...) is defined as the Shannon entropy for $k$ distinct symbols:

$$H_p(n) = -\sum_{j=1}^{k} P_j \ln(P_j)$$ (12)

When $P_j = \frac{1}{n!}$, then $H_p(n)$ attain the maximum value $\ln(n!)$ [3,5]. For convenience, we always normalized:

$$0 \le \frac{H_p(n)}{\ln(n!)} \le 1$$ (13)

We can define the permutation entropy per symbol[5,15] of order $n$

$$h_n = \frac{H(n)}{n-1}$$ (14)

In our calculation the order of permutation entropy is taken *n*=4.

### 2.3 . Multiscaling Entropy

Another approach for investigation of complexity of biological signal is multiscaling entropy. Given a one-dimensional discrete time series, $\{x_1,...,x_i,...,x_N\}$, Madalena et.al. construct consecutive coarse-grained time series, $\{y^{(\tau)}\}$ determined by the scale factor $\tau$ according to the equation $y_j^{(\tau)} = \frac{1}{\tau} \sum_{i=(j-1)\tau+1}^{j\tau} x_i$. For scale one, the time series $\{y^{(1)}\}$ is simply the original time series. The length of each coarse-grained time series is equal to the length of the original time series divided by the scale factor $\tau$. Madalena et.al. then calculate an entropy measure for each coarse grained time series plotted as a function of the scale factor $\tau$ [16-18]. Madalena et.al. called this procedure multiscale entropy (MSE) analysis[17].

## 3. Result and discussion

### 3.1. Equilibrium time of viral kinetics model as a function of template number:

Kinetics Monte Carlo approach is used for simulation of intracellular viral reaction kinetics via Gillespie algorithm. Stochastic dynamics for investigation number of template particle is investigated as a function of time. Equilibrium time after 1000 times of stochastic simulation is obtained as a function of initial template population.

On the basis of Figure 1 we find the scaling relation between the equilibrium of time and the number of template particle. Fitting of Figure 1 with $f_{eq\,time}(N) = aN^b$ shows $a$ = 163.1 , $b$ = -0.1429 and R square = 0.9844. Kinetics Monte Carlo shows that at Figure 1 with increasing the number of template, time of equilibrium decreases as a power law.

Relation between the time of equilibrium $f_{eq\,time}(N) = aN^b$ shows that time decreases very fast when number of template particle increases. Figure 1 shows that there is small time for equilibrium time for large number of template species. Increasing the number of template decreases the fluctuation population of template, then system approaches to stationary sate at the short time; therefore, system approaches to equilibrium in short time. Low population of template and existence of nonlinear dynamics, reaction in step five, cause more fluctuation; therefore; there is large value for equilibrium time in low population of template.

## 3.2. Permutation entropy Result

Stochastic simulations have been done for intracellular viral reaction kinetics via Gillespie algorithm. Population of structural protein, genome and template has been calculated as a function of time. The result of stochastic simulation for one run for population dynamics of genome, structural protein, and template with two kinds of algorithm Gillespie, Poisson are shown in Figure 2 a,b respectively. Permutation entropy base on comparison neighborhood value equations 12, 14 has been calculated from the signal which obtained from stochastic simulation of intracellular viral reaction kinetics. Stochastic simulation has been done for 1000 independent runs for genome, template and structural protein and in each run permutation entropy has been calculated, therefore average of 1000 permutation entropy in each time is computed. As a result, average of 1000 independent permutation entropy in during 200 days for genom species for four different template populations, say low and high infection, is shown in Figure4.a. On the basis of Figure.4.a in low simulation time permutation entropy of genome in low-high infection level , say less than 5 template population number, is smaller than very high infection level however, permutation entropy trend is reversed after more  5 template population number in big time in order permutation n=4. For big permutation entropy order n=6 permutation entropy increases with the initial template population monotonically.   Result of permutation entropy from averaging of 1000 independent permutation entropy from kinetics Monte Carlo population trajectory as a function of time for structural protein with initial population number Temp = 2,3 and

 Temp =5-7 are presented at Figure.4.b and Figure.4.c respectively. On the basis of Figure.4.b in low infection level of intracellular viral reaction kinetics, there are two peaks as a sharp and wide in short time scale, finally permutation entropy for structural protein decreases in large time scale linearly. For very high infection level, permutation entropy of structural protein, just one peak observed in small time scale and permutation entropy trend is linear in big time scale. Permutation entropy has been calculated for template dynamics in low and very high infection level. Result of permutation entropy of template has been presented at Figure.4.d. On the basis of Figure.4.d.  permutation entropy of template increases with infection level.

Result of permutation entropy as a function of time with three permutation order say, n=2, 4, 6 with low and very high infection level has been shown in Figure.5.a,b respectively. On the basis of permutation entropy at Figure.5.a and Figure5.b, increase of infection level in intracellular viral kinetics lead to convergency of permutation entropy with the low permutation order (n). Convergency of permutation entropy has been tested on windows value and permutation order (n ) value. Dependency of permutation entropy on order and windows value at (n=4, windows=512) and (n=6, windows=1024) are presented at Figure.5.c and Figure.5.d respectively. On the basis of Figure.5.c and

Figure.5.d the value of windows at high permutation order (n=6) on permutation entropy value is disappeared completely. Permutation entropy convergency occurs on six permutation order value. Permutation entropy calculation shows that at short time scale in intracellular reaction dynamics convergency occurs with medium value permutation order but by passing time dynamical entropy depend on n value and in big time scale permutation entropy shows as a scaling law $H(n) = n^{\alpha}$ ($\alpha$=0.30). Fitting result for permutation entropy per symbol $h(n)$ has been presented at Figure.5.e.

Comparison of permutation entropy for three kinds of species such as structural protein, template and genome from averaging of entropy production based on five initial template number and 1000 stochastic simulation runs are presented at Figure.6.a and Figure.6.b respectively. On the basis of Figures.6 permutation entropy versus time (day) has a following order: template $>$ structural protein $>$ genome.

Noise effect has been investigated on permutation entropy for stochastic dynamics of intracellular reaction kinetics. Permutation entropy result in absence and presence of noise for dynamical investigation of intracellular reaction in very high infection effect is shown in Figure.6c and Figure.6.d respectively. On the basis of Figure.6.c and Figure.6.d presence and absence of noise including dynamics on magnitude of permutation entropy are not same to each other. Owing to this fact noise can change the entropy value and magnitude order entropy for genome, template and structural protein but including noise in permutation entropy is not able to predict the complexity of stochastic dynamics of intracellular viral reaction.

Figure 6.c indicates that permutation entropy for template species is greater than genome species. On the basis of Figure.6.c permutation entropy for genome species is bigger than structural protein in intracellular viral model which may be doubtable prediction. According to Figures 2.a,b and Figure 3 dynamical trajectory for the population of template is unpredicted than the other two species namely genome and structural proteins species. The trend of permutation entropy is not flat for three species namely structural protein, genome, and template. More permutation entropy shows that fluctuation population of template as a function of time is more than genome and structural species. On the basis of Figure 2 stochastic result of genome shows that population of genome is not so sharp, and it does not have a large fluctuation population as a function of time. As a result, permutation entropy is not able to give correct result regarding the complexity of stochastic dynamics of intracellular viral reaction.

### 3.3. Multi Scaling Entropy

### 3.3.1. Multi scaling entropy with including constant standard deviation of population

### 3.3.1.a. Multi scaling entropy for genome

For the comparison of complexity of the dynamic and multiscaling entropy production trend in intracellular viral reaction kinetics we continue our calculation on the basis of multi scaling entropy. Gillespie algorithm is used for stochastic simulation of the mentioned model. The result of time evolution for population from stochastic run is used for computation of multiscaling entropy. We make coarse graining of population number of each species from simulation data. Multiscaling entropy is calculated on the basis reference [16-18].

Let $\{X_i\} = \{x_1, x_2, ..., x_N\}$ represent a time series of population each species of length $N$ in stochastic dynamics of intracellular reaction kinetics. Consider the $m$-length vectors:

$u_m(i) = \{x_i, x_{i+1}, \ldots, x_{i+m-1}\}$ following definition for the distance between two vectors in intracellular reaction dynamics

$d[u_m(i), u_m(j)] = \max[|x(i+k) - x(j+k)| : 0 \le k \le m-1]$ Let $n_{i,m}(r)$ represent the number of vectors $u_m(j)$ within $r$ of $u_m(i)$, Therefore; $C_i^m(r) = \dfrac{n_{im}(r)}{N-m+1}$ represents the probability that any vector $u_m(j)$ is within $r$ of $u_m(j)$. Define

$$\varphi^m(r) = \frac{1}{(N-m+1)} \sum_{i=1}^{N-m+1} \ln C_i^m(r) \qquad (18)$$

For calculation of multiscaling entropy, we calculated standard deviation(SD) our initial simulation data then, we compute number of data which set in distance r = 0.15SD. The result of multiscaling entropy as a function of scaling factor for low to very high infection level for genome species is shown in Figure 7.a. According to the Figure.7.a multi scaling entropy value increases with template in range number 1-5 and after 5 templates as a initial population for intracellular reaction dynamics, multi scaling decreases with template significantly. On the basis of Figure.7.a multi scaling entropy increases with scale factor linearly.

### 3.3.1.b. Multi scaling entropy for structural protein

Multi scaling entropy calculation has been extended for structural protein versus scaling factor. Multi scaling entropy result for structural protein shows that there is linear relation relationship versus scaling factor in initial population of template at range 1-8. Increase of initial population from 9-10 for template shows one peak in multi scaling entropy as a function of scaling factor for structural protein. Result of multiscaling entropy f structural protein at initial template population (2,3) and 10 numbers are presented at Figure.7.b and Figure7.c respectively.

### 3.3.1.c. Multi scaling entropy for Template

Multi scaling entropy has been calculated for low, medium and very high infection level of template versus scaling factor. For low, medium, and high infection level multi scaling entropy trend is completely different to each other. Result of multi scaling entropy versus scaling factor for low, medium and high infection level for template dynamics has been presented at Figure 8.a, b, c respectively. On the basis of Figure.8.a in low infection level multiscaling entropy increases with scale factor linearly. According to the Figure.8.a multiscale entropy for template species versus scale factor increases in low infection level of template. In medium infection level, there is non monotonic behavior for multiscaling entropy versus scaling factor. Result of permutation entropy as a function of scaling factor in medium infection level has been presented at Figure.8.b. Multiscaling entropy trend in high infection level for stochastic dynamics in intracellular reaction is reversed in comparison with low infection level completely. For high infection level multiscaling entropy decrease with the scaling factor and there is crossover for multiscaling entropy versus scaling factor with the different initial population of template. Result of multiscaling entropy as a function of scaling factor in high infection level is presented at Figure.8.c.

Comparison of multiscaling entropy versus scaling factor for intermediate infection level of template for template, structural protein and genome species has been done and its result has been presented at Figure.8.d. On the basis of Figure.8.d there is a following order for multi scaling entropy: template> structural protein > genome.

### 3.3.2. Scaling entropy with including variation standard deviation of population

For computation of multiscaling entropy in part 3.3.1 standard deviation was constant and $r$ parameter was set $r = 0.15\text{SD}$. As a matter of fact for each scaling factor in during of coarse graining method standard deviation is not constant.[18] Multiscaling entropy on the basis variation of standard deviation (modified multi scaling entropy) from Gillespie algorithm for stochastic dynamics is calculated for each scaling factor parameter. For each scaling factor standard deviation is calculated; therefore, $r$ is set as $0.15\text{SD}$. On the basis of the mention principal multiscaling entropy (modified multiscaling entropy) is calculated as a function of scaling factor. Computation result of modified multiscaling entropy for genome species for each scaling factor is shown in Figure 9.a. On the basis of Figure.9.a modified multiscaling entropy changes with scaling factor linearly for all initial template population. Modified multiscaling entropy value increases with low-high infection level (1-5) and for more than 5 initial template population, multiscaling entropy decreases again.

Modified multiscaling entropy of structural protein shows that modified multiscaling entropy increases with scaling factor linearly in the initial template range (1-8). Result of modified multi scaling entropy for structural protein with one initial temple population has been presented at Figure.9.b. When number of initial template increases to 10, then one maximum peak observed in modified multiscaling entropy for structural protein.

Modified multi scaling entropy calculation has been extended for template species for different initial population of template and its result is presented at Figure.9.d. On the basis of Figure.9.d modified multi scaling entropy decreases with scaling factor monotonically. It is worthwhile to notice that there is no regular trend for multi scaling entropy versus scaling factor for template species; however modified multiscaling entropy of template has a regular trend for different initial number of template.

Comparison of modified multiscaling entropy versus scaling factor has been done for three kinds of species, namely genome, structural protein and template. Result of comparison for modified multiscaling entropy has been shown in Figure.9.e. On the basis of Figure.9.e there is a following order for modified multiscaling entropy

Template> Structural Protein > genome.

### 3.3.3. Initial Template effect on permutation entropy, multiscaling entropy and modified multiscaling entropy

Permutation entropy for three kind of species namely genome, structural protein, template has been calculated for different initial template population. On the basis of permutation entropy result, there is monotonic behavior for permutation entropy value versus initial population of template in all range of time. Calculation of multiscaling entropy has been extended as a function of initial template molecule. Multiscaling entropy shows that there is a non-monotonic behavior for entropy versus initial template population for all species namely gen, structural protein and template. Modified multiscaling entropy calculation has been done in different infection level and non-monotonic behavior has been found for modified multiscale entropy for genome, structural protein and template as well. Result permutation entropy, multiscaling entropy and modified multiscaling entropy versus initial template molecule for template species is presented at Figure.10.a, b, c respectively.

## 4. Conclusion

Kinetics Monte Carlo simulation has been done for intracellular viral reaction kinetics. Gillespie algorithm is used for investigation the population dynamics of three species structural proteins, template, and genome. Equilibrium time of template is obtained as a function number of template species via stochastic simulation by using Gillespie algorithm. On the basis of kinetics Monte Carlo result, there is power law as a formula $f_{eq\ time}(N) = aN^b$ between equilibrium time and initial population of template species. Stochastic simulation shows that equilibrium time decreases with initial number of template molecule. Permutation entropy on the basis of comparison neighborhood value is calculated for three species of viral reaction kinetics. Permutation entropy result depends on windows in low permutation order, however in high permutation order n >=6 the permutation entropy value is not depend on the windows value. Permutation entropy in presence chaotic noise has the following order template> genome> structural protein. In absence of noise permutation entropy value has a different order such as structural protein > genome>template. Two clear peaks for permutation entropy of structural protein can be observed in low infection level however in high infection level one peak is disappeared completely and just one peak is observed versus time (day). In addition to multiscaling entropy is calculated on the basis of constant and variation standard deviation for each scale factor. Multiscaling entropy and modified multiscaling entropy have same prediction regarding the magnitude of entropy value and they show the following order: template> structural protein> genome. It is worthwhile to notice that the result of trend for multiscaling entropy of template versus scaling factor depends on initial template population. There is three different behaviors for multiscaling entropy for template versus scaling factor. In low infection level mutiscaling entropy increases with scaling factor monotonically however in high infection level there is minimum in mutiscaling entropy. In very high infection level multiscaling entropy decreases versus scaling factor monotonically. Modified multiscaling entropy for template species decreases with scaling factor for all initial template population monotonically. For genome species, multiscaling entropy and modified multiscaling entropy versus scaling factor shows monotonic behavior. For structural protein multiscaling entropy and modified multiscaling entropy in low and medium infection level versus scaling factor is monotonic however in very high infection level three is a peak for both multi and modified multiscaling entropy as a function of scaling factor. Permutation entropy in presence and absence of chaotic noise is not able to predict complexity of intracellular viral kinetics dynamics however multiscaling and modified multiscaling entropy estimate the natural complexity of intracellular viral kinetics dynamics.

**Figure Captions:**

Figure.1. Equilibrium time for template species as a function of initial population of template.

Figure.2. Typical one kinetics Monte Caro for genome species with two different kinds of algorithms, Gillespie, Poisson. b) Kinetics Monte Carlo simulation for structural protein as a function of time via Gillespie, Poisson algorithm.

Figure.3. Same as Fig.2 one kinetics Monte Carlo simulation run for template with Gillespie and Poisson algorithms.

Figure.4. a) Permutation entropy of genome as a function of time for initial template with number 1, 5,6,10. b ) Permutation entropy as a function of time for structural protein at initial template molecule with population 2 and 3. c) Similar to the Figure.4.b permutation entropy for structural protein with the initial population template number 5,6,7,10 d.

d) Permutation entropy of template as a function of time (day) with initial population of template 1,5,10

Figure.5. a) Permutation entropy of genome versus time (day) for different order of permutation entropy n=2, 4, 6 for initial template=1. b) Similar to Figure.5.a permutation entropy for different order n=2, 4, 6 with initial template=10.

c) Permutation entropy for genome versus time for order n=4 but for different windows =512, 1024.

d) Similar to the Figure.5 permutation entropy for genome as a function of time at windows 512, 1024 with n=6.

e) Behavior of permutation entropy per symbol h(n) versus n in big time scale of dynamical system.

Figure.6. a) Comparison permutation entropy on the basis of comparison neighborhood value for three species, genome, structural protein, template as a function of time (day) from result of one  stochastic trajectory of population with initial template 5.

b) Similar to the Figure.6.a comparison of permutation entropy for three species, genome, structural protein from result of 1000 stochastic population trajectory.

c) Comparison of permutation entropy for genome, structural protein, template with initial template=10 in absence of noise

d) Comparison of permutation entropy for genome, structural protein, template with initial template=10 in presence of noise

Figure.7.a) Multiscaling entropy for genome species versus scaling factor with initial template with number 2, 5,6,10.

b) Multiscaling entropy for structural protein versus scaling factor with initial template 2,3.

c) Multiscaling entropy for structural protein versus scaling factor with initial template 10

Figure.8. a) Multiscaling entropy for template as a function of scaling factor for initial template 1,3.

b) Multiscaling entropy for template versus scaling factor for initial template=4

c) Same as Figure.8.b, Multiscaling entropy for template versus scaling factor for initial template=7,10

d) Comparison of multiscaling entropy for template, genome, structural protein versus scaling factor with initial population of template=3

Figure.9. a) Modified scaling entropy of genome versus scaling factor with different initial template population 1,3,5,6,10

b) Modified scaling entropy of structural protein versus scaling factor with initial template population 1.

c) Same as Figure.9.b modified scaling entropy of structural protein versus scaling factor for initial template=10

d) Modified scaling entropy of template as a function of scaling factor with initial template molecule 1,3,4,5,6,10.

e) Comparison modified multi scaling entropy genome, structural protein, template versus scaling factor with initial template=3.

Figure.10. a) Permutation entropy of template versus initial template population

b) Multiscaling entropy of template versus initial template population

c) Modified multiscaling entropy of template versus initial template population

**Figure 1**

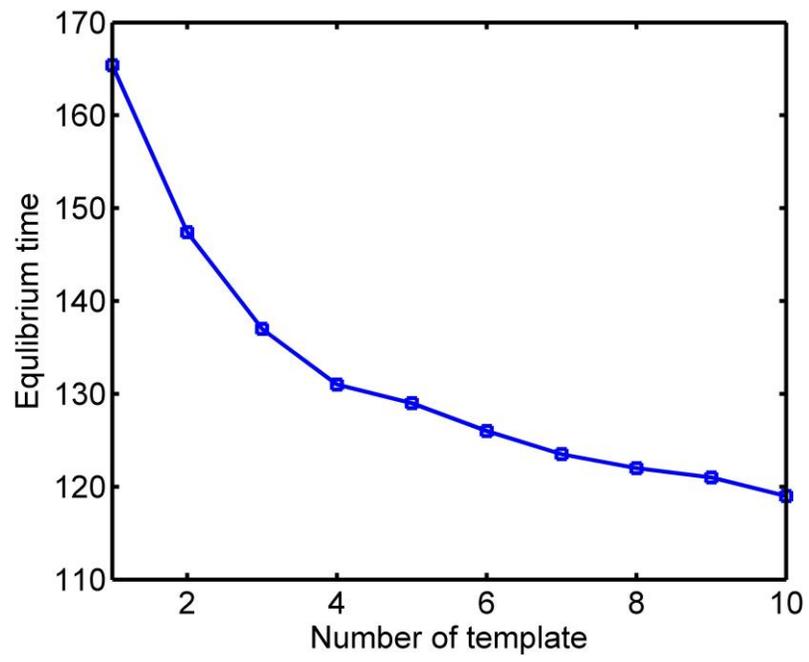

**Figure.2.a**

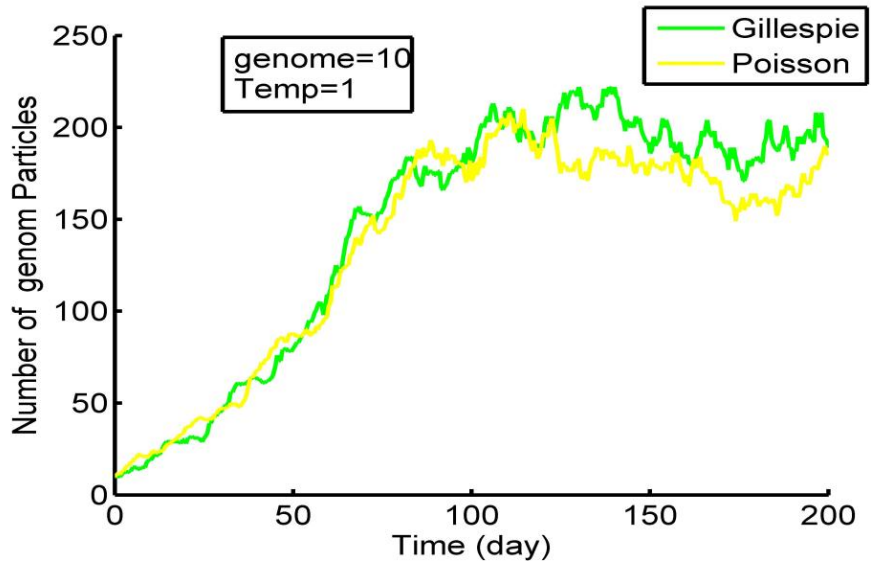

**Fig.2.b**

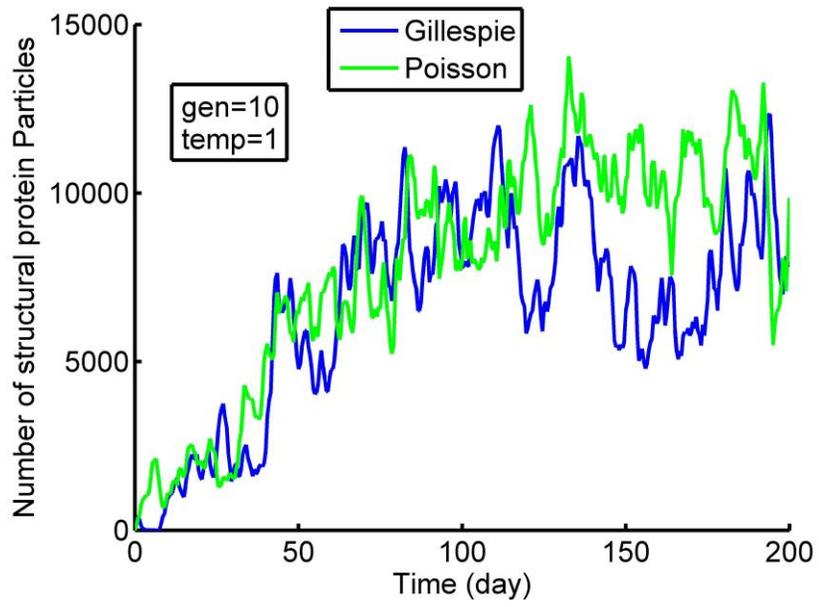

**Figure.3.**

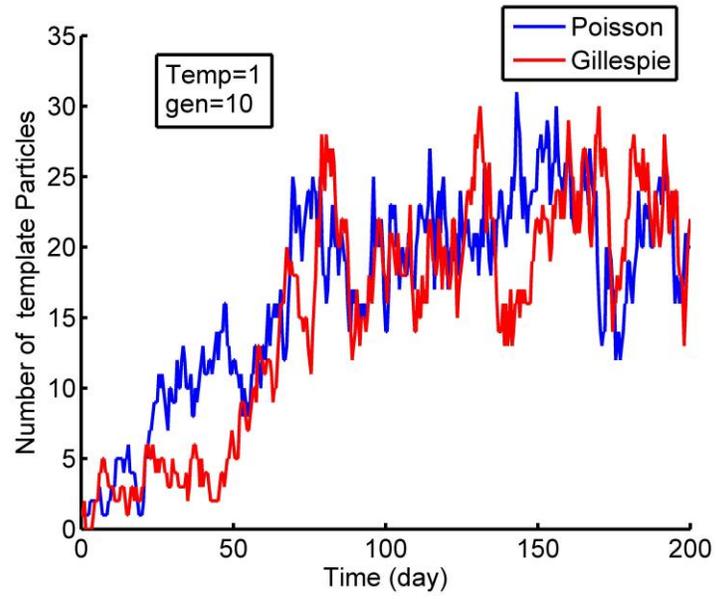

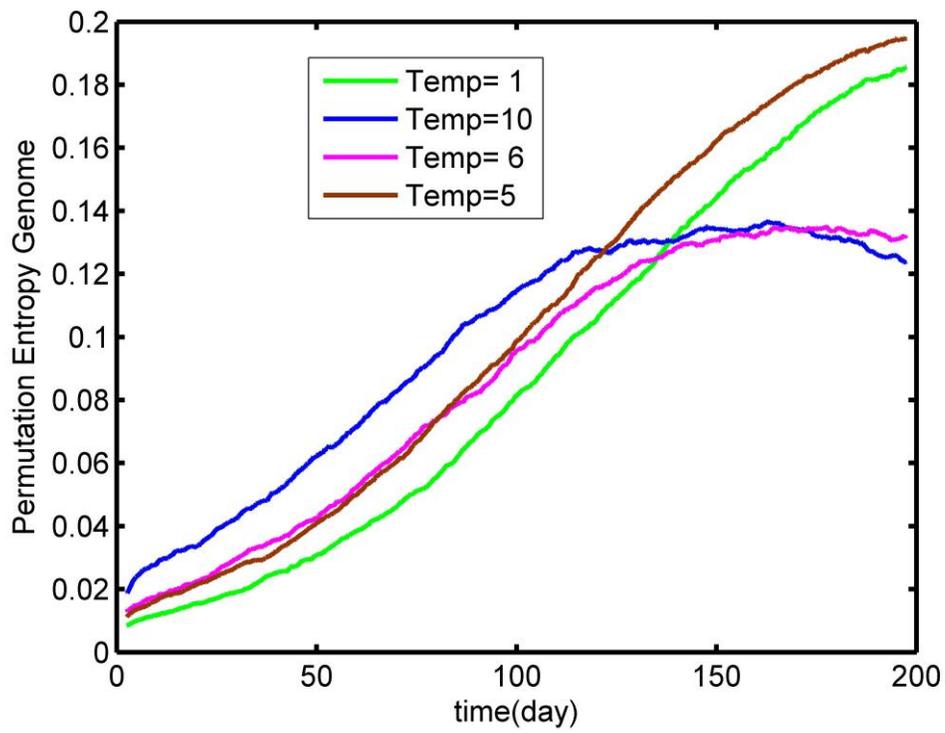

Figure.4.a

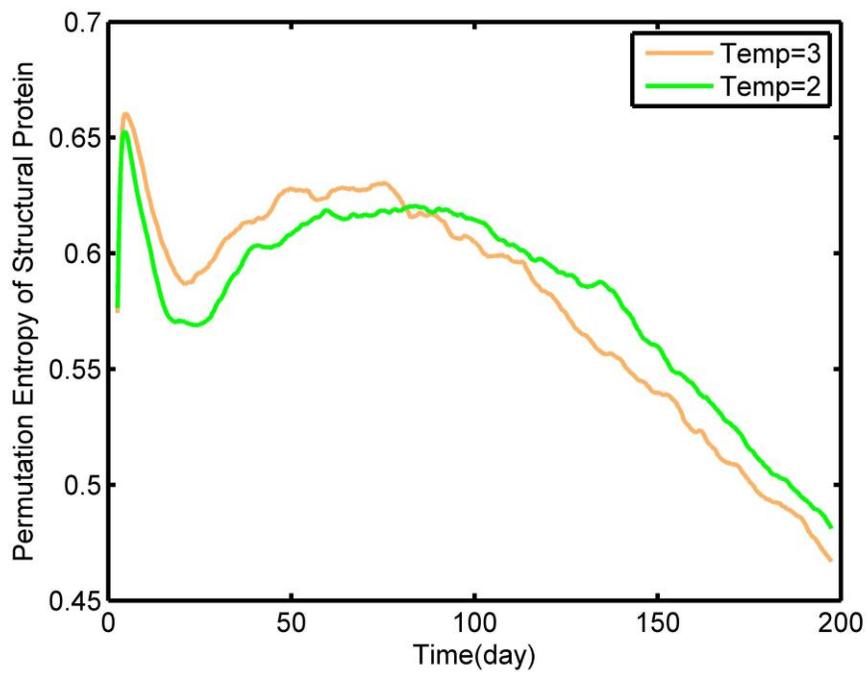

Figure.4.b

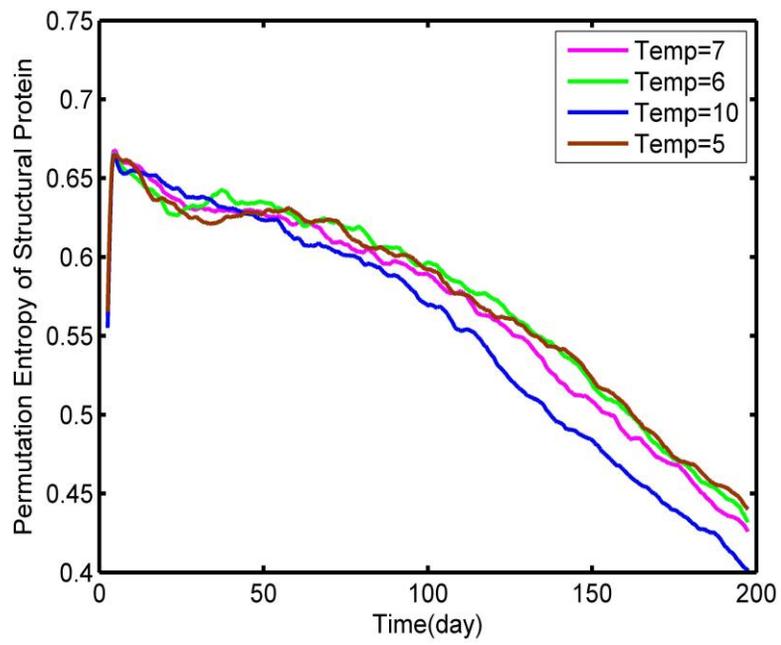

Figure.4.c

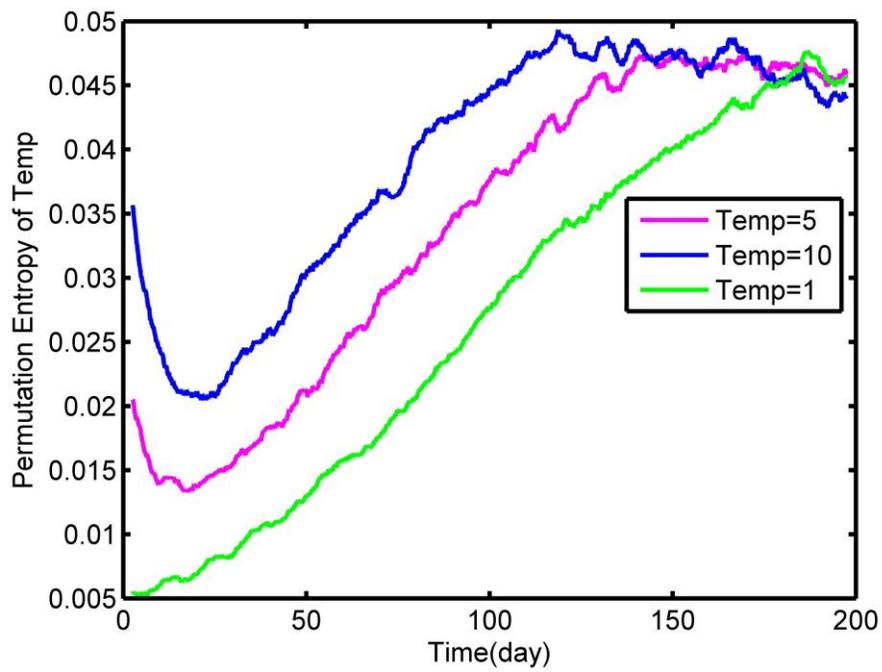

Figure.4.d

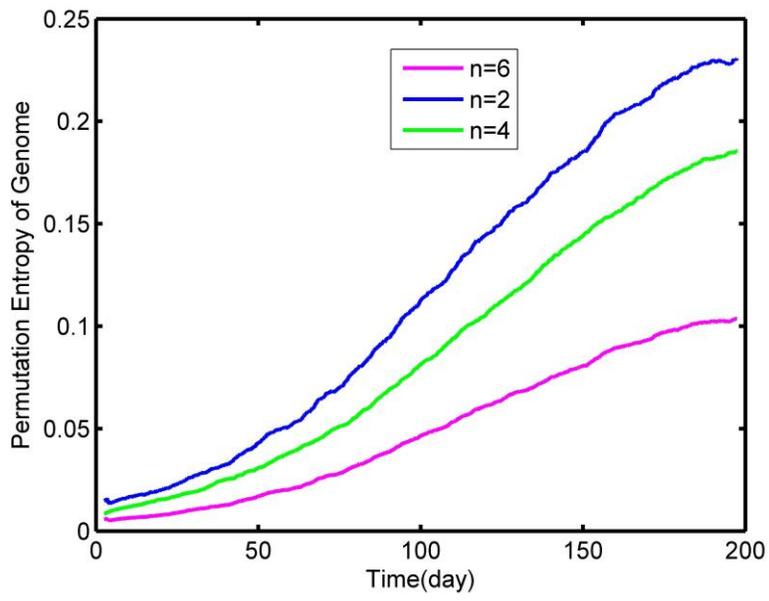

Figure.5.a

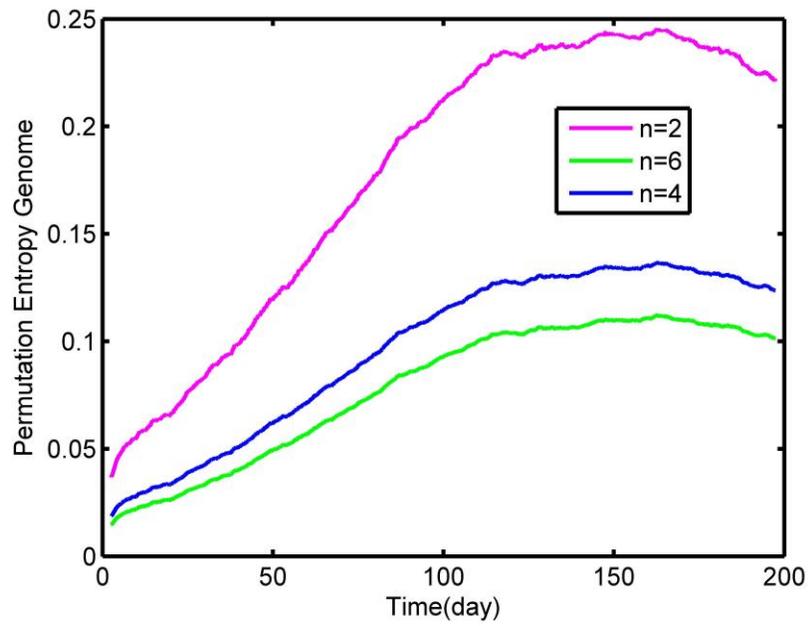

Figure.5.b

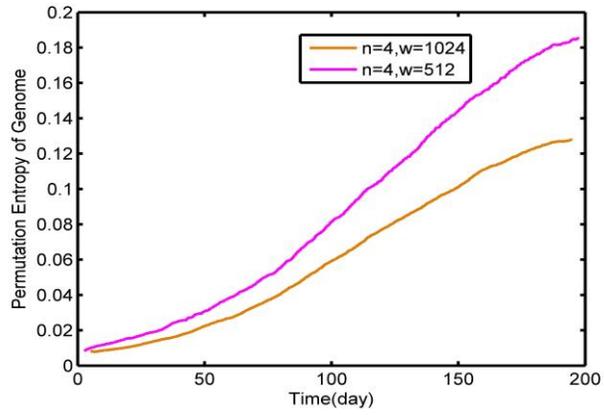

Figure.5.c

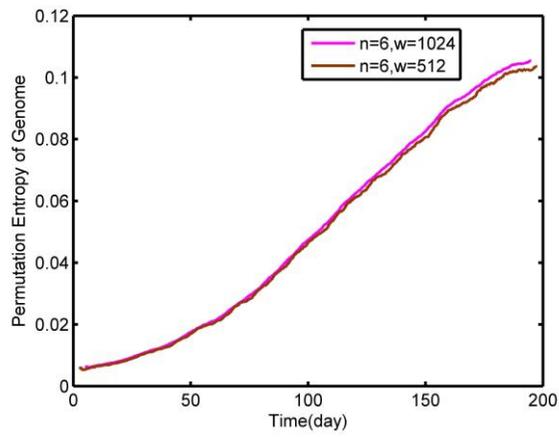

Figure.5.d

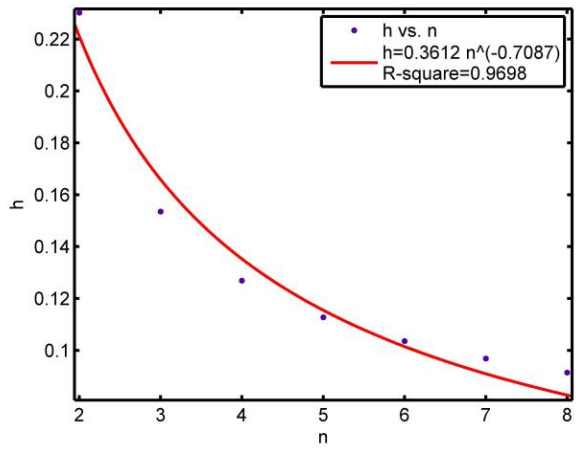

Figure.5.e

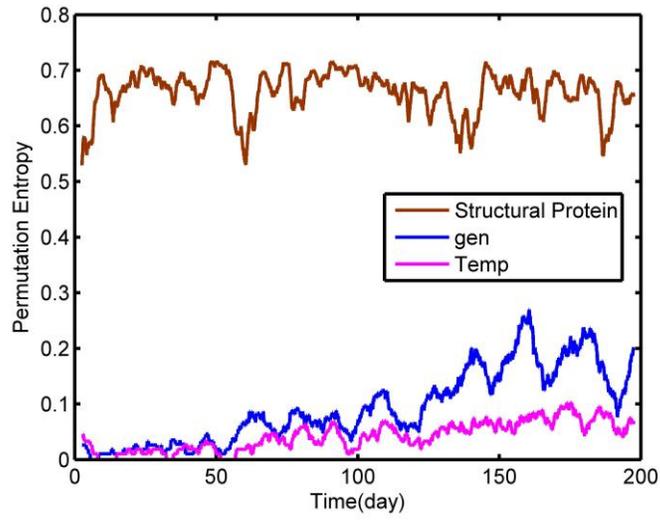

Figure.6.a

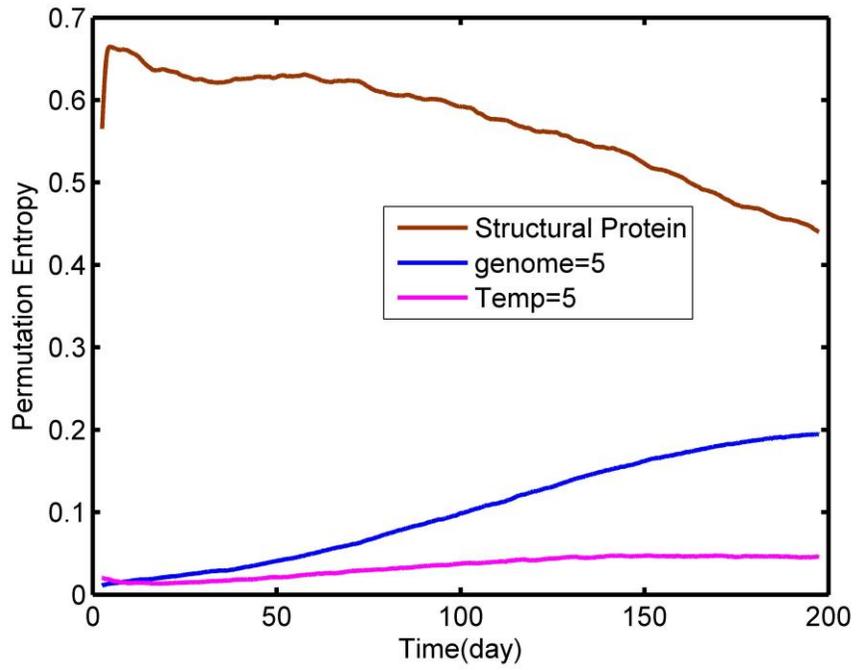

Figure.6.b

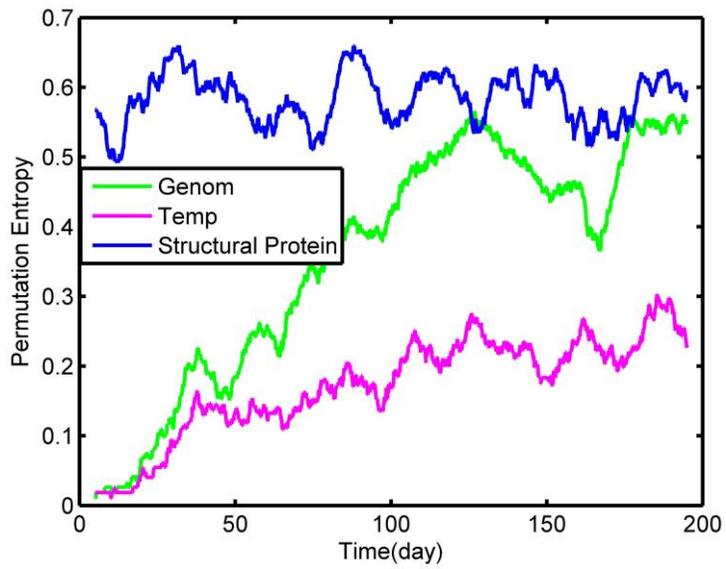

Figure.6.c

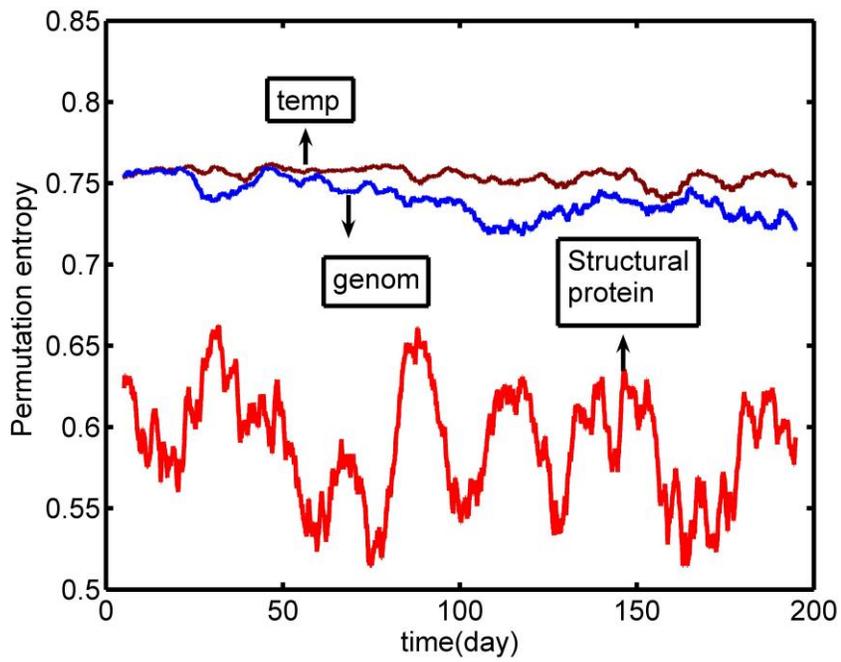

Figure.6.d

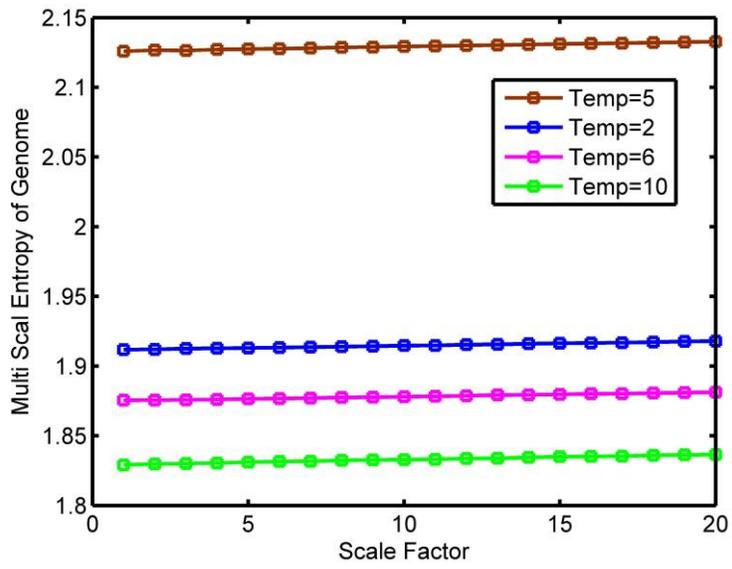

Figure.7.a

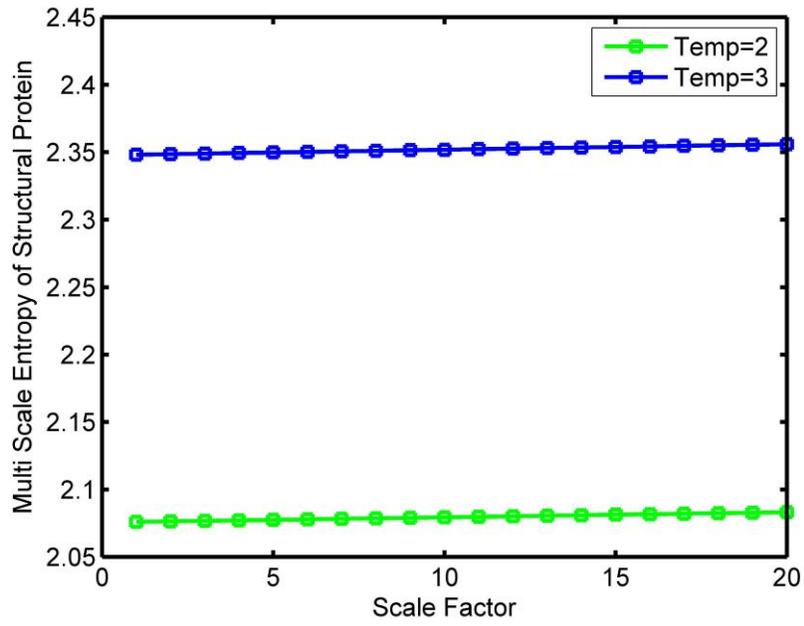

Figure.7.b

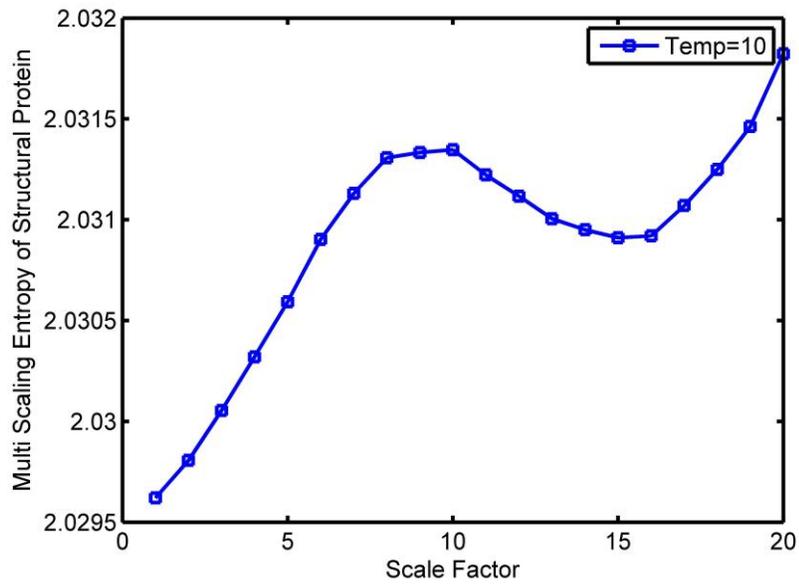

Figure.7.c

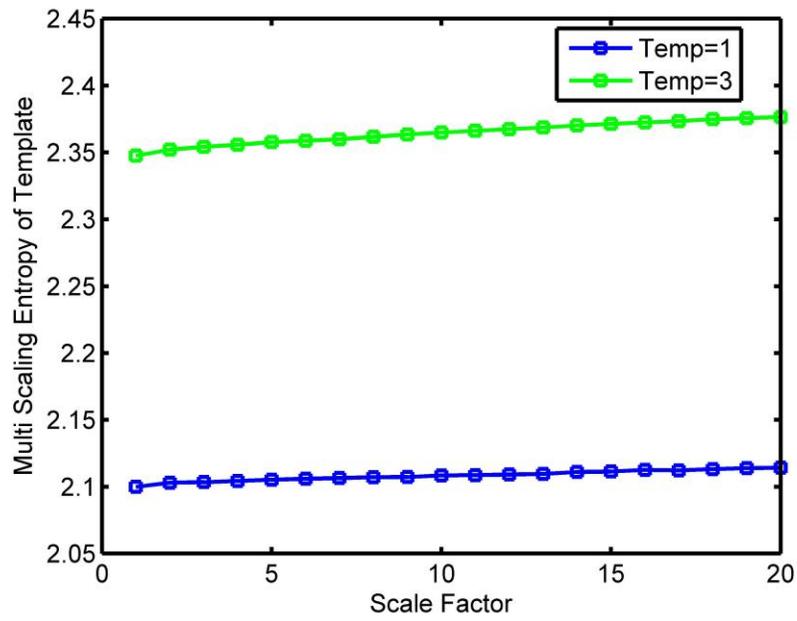

Figure.8.a

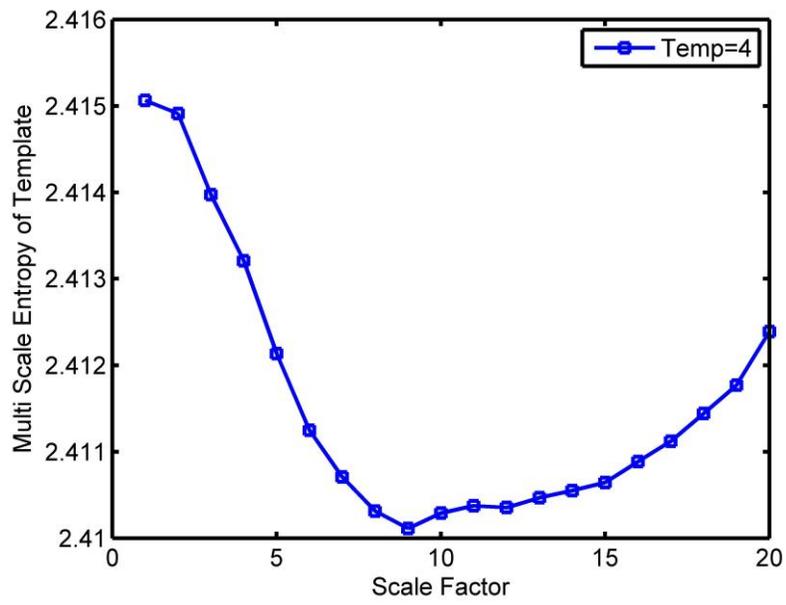

Figure.8.b

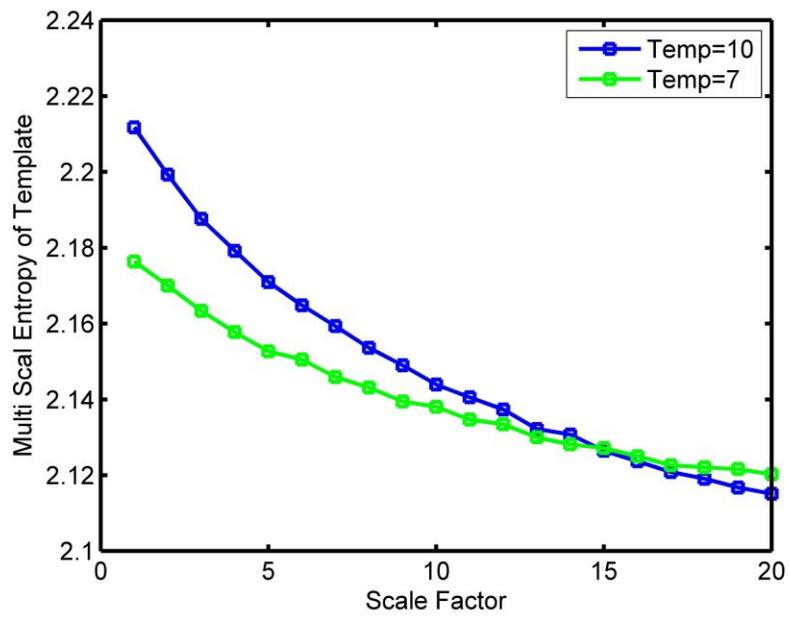

Figure.8.c

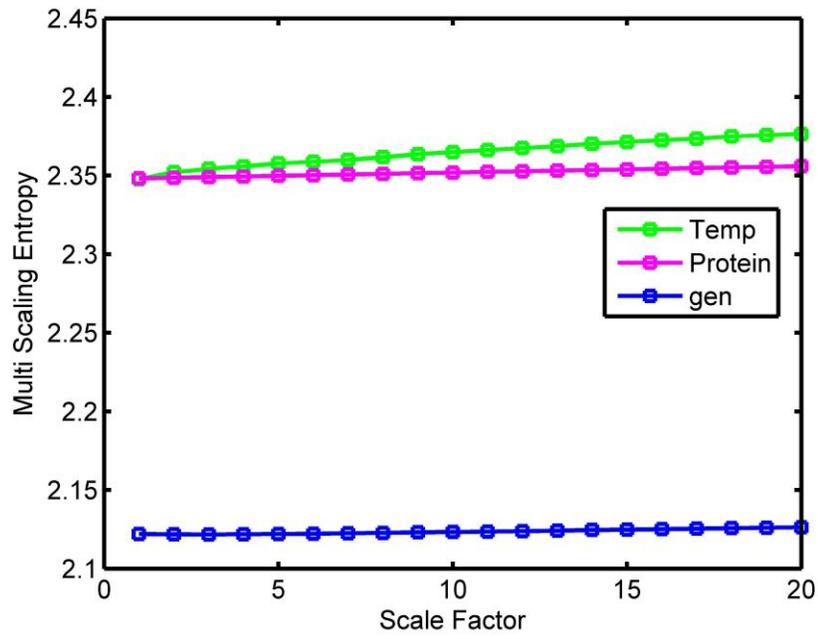

Figure.8.d

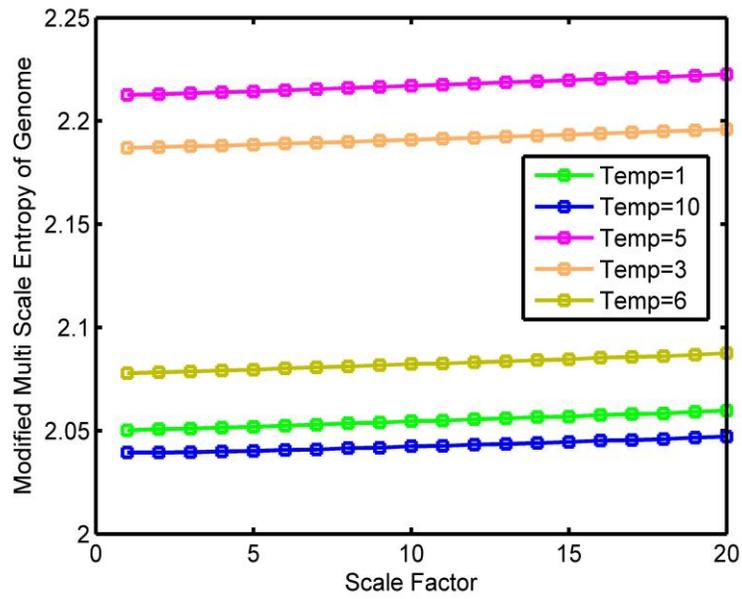

Figure.9.a

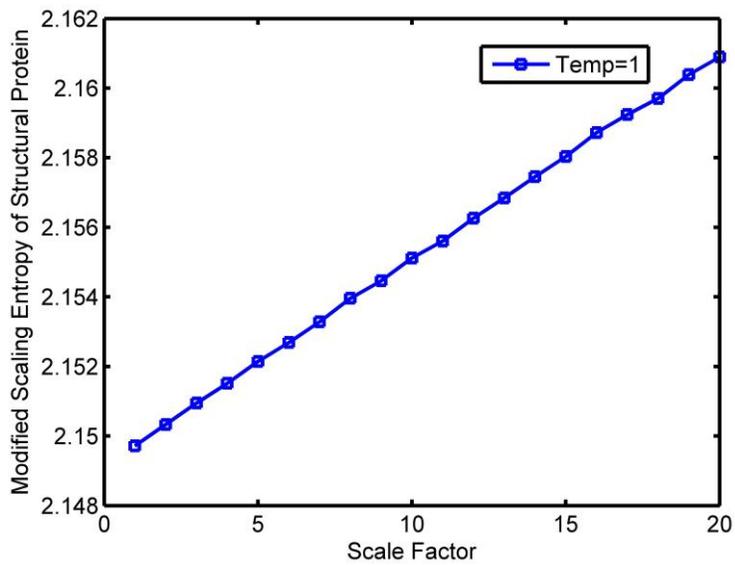

Figure.9.b

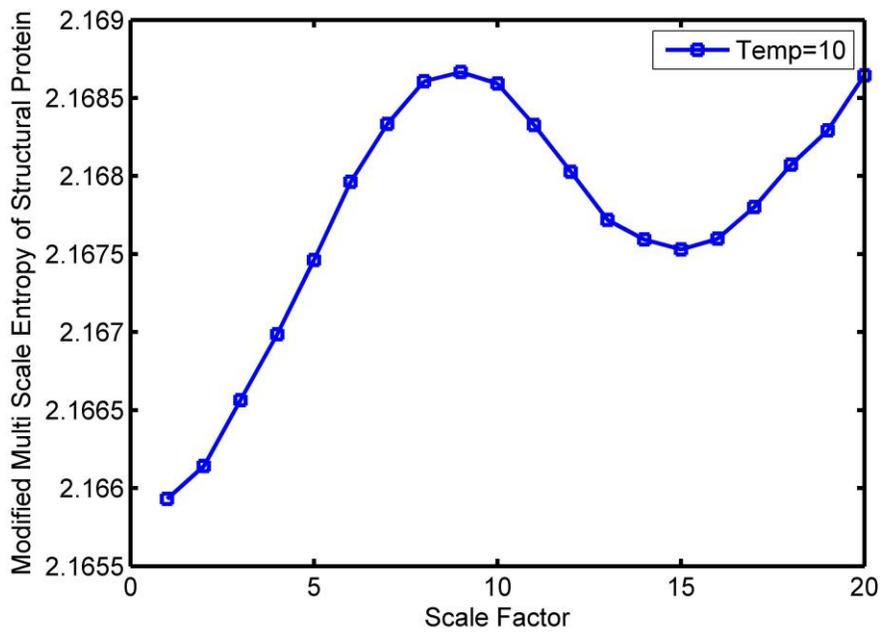

Figure.9.c

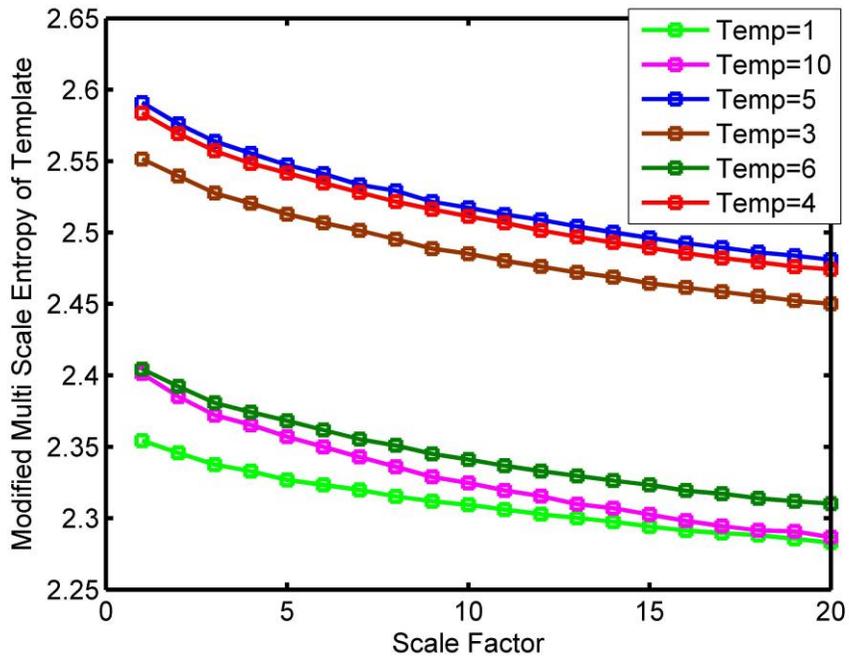

Figure.9.d

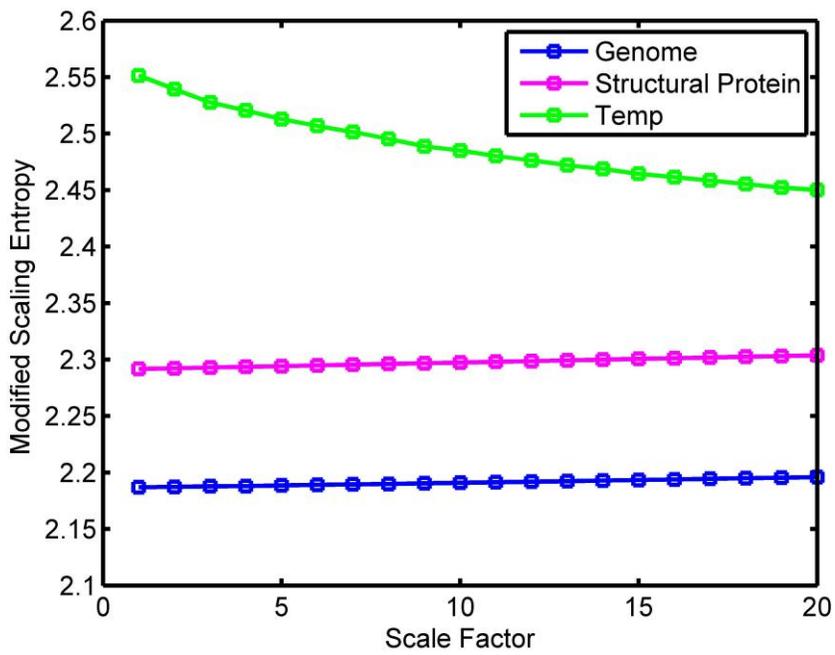

Figure.9.e

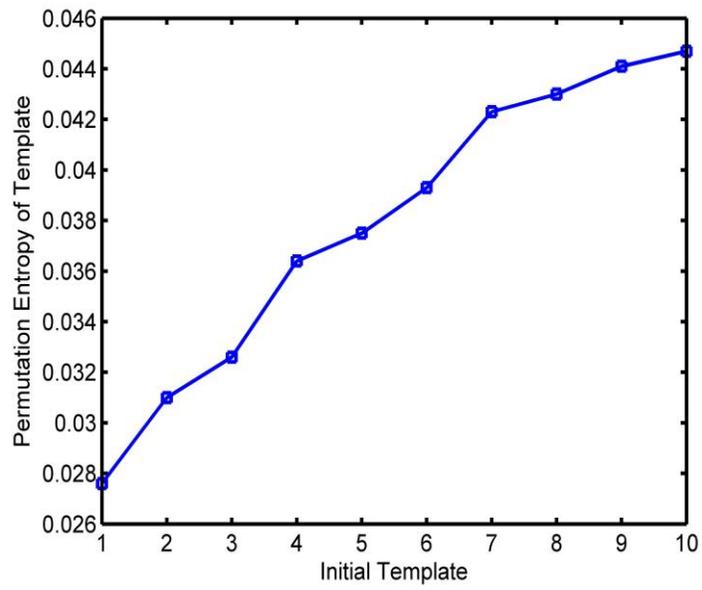

Figure.10.a

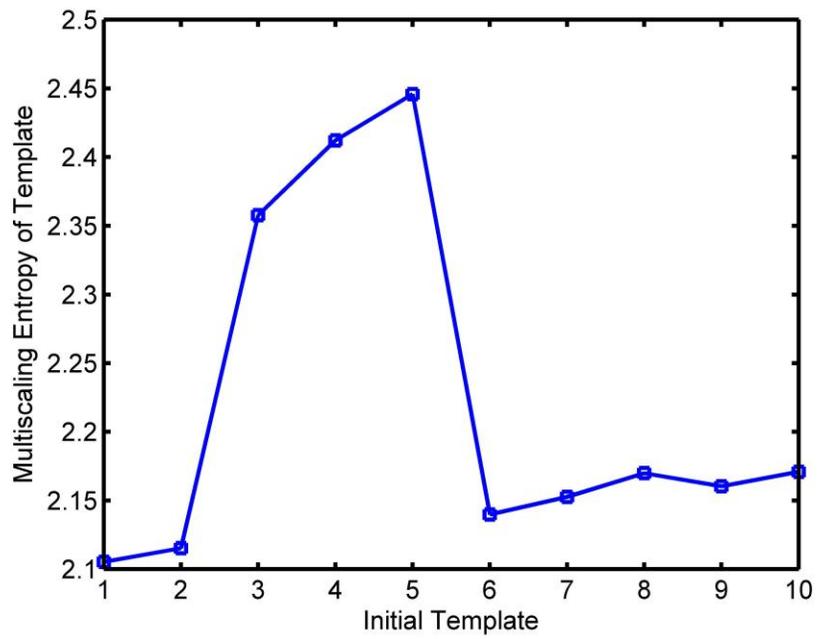

Figure.10.b

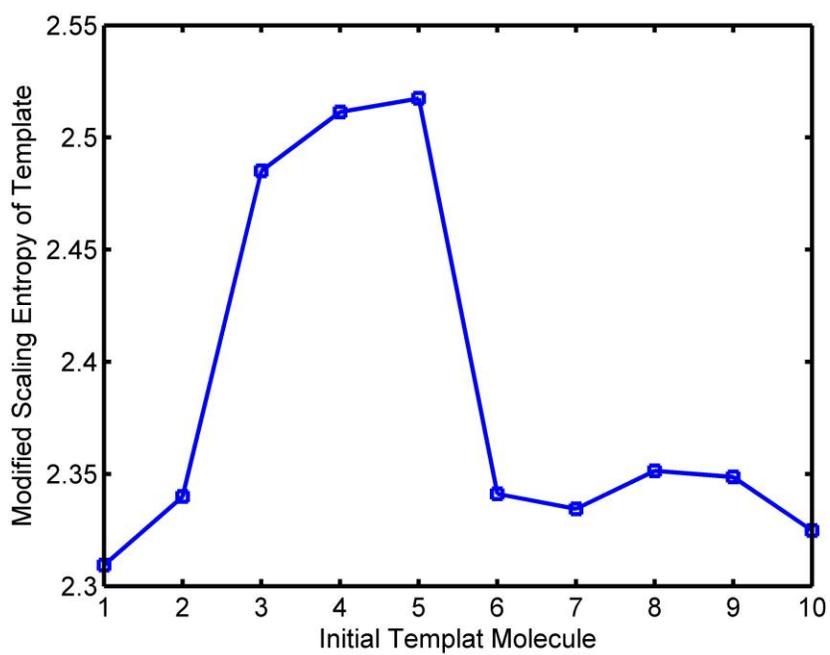

Figure.10.c